\documentclass[english,prb,longbibliography,floatfix,superscriptaddress,twocolumn,showpacs,amsmath,amssymb,reprint,nofootinbib,latin9]{revtex4-2}
\usepackage[T1]{fontenc}
\usepackage[latin9]{inputenc}
\usepackage{color}
\usepackage{amsmath}
\usepackage{amssymb}
\usepackage{graphicx}
\usepackage{float}
\usepackage{bbm}
\usepackage[normalem]{ulem}

\makeatletter

\newcommand{\lyxmathsym}[1]{\ifmmode\begingroup\def\b@ld{bold}
  \text{\ifx\math@version\b@ld\bfseries\fi#1}\endgroup\else#1\fi}

\providecommand{\tabularnewline}{\\}

\PassOptionsToPackage{caption=false}{subfig} 
\usepackage{hyperref}
\hypersetup{
breaklinks=true,
colorlinks=true,
citecolor=blue,
linkcolor=blue,
filecolor=blue,
urlcolor=blue
}
\IfFileExists{lmodern.sty}{\usepackage{lmodern}}{}

\makeatother

\usepackage{babel}
\usepackage{xcolor}

\begin{document}

\title{$G_2$ Integrable Point Characterization
via Isotropic Spin-3 Chains}

\author{Chengshu Li}%
 \email{lichengshu272@gmail.com}
 \affiliation{Institute for Advanced Study, Tsinghua University, Beijing 100084, China}
 
 \author{Victor L. Quito}
 \email{vquito@iastate.edu}
\affiliation{Department of Physics and Astronomy, Iowa State University, Ames, Iowa 50011, USA}
\affiliation{Ames National Laboratory, Ames, Iowa 50011, USA}

\author{Dirk Schuricht}%
\email{d.schuricht@uu.nl}
\affiliation{Institute for Theoretical Physics, Center for Extreme Matter and Emergent Phenomena, Utrecht University, Princetonplein 5, 3584 CE Utrecht, The Netherlands}

\author{Pedro L. S. Lopes}
 \email{plopes@quera.com}
\affiliation{QuEra Computing Inc., 1284 Soldiers Field Road, Boston, MA 02135, USA}

\begin{abstract}
We investigate the physical properties of $G_2$-symmetric integrable chains with local degrees of freedom in the fundamental representation; given the typical connection between integrability and critical points, we test the model's properties against a hypothesis of conformal-invariant long-distance behavior. Leveraging an embedding between the  $G_2$ exceptional Lie algebra and $SU(2)$-symmetric chains with local spin-3 representations, we perform numerical analyses via  exact diagonalization (ED) targeted at specific spin sectors, as well as via non-Abelian density-matrix renormalization group (DMRG). A basic study of the momentum-resolved ED spectrum suggests the low-energy system is effectively described by a $(G_2)_1$ Wess--Zumino--Witten (WZW) theory, but we find challenges in further numerical characterization of conformal data. The study and control of the phenomenology of this model may have implications for the development of accessible models for Fibonacci anyons.
\end{abstract}

\date{\today}

\maketitle

\section{Introduction}
Exceptional Lie algebras bring exotic and rich emergent phenomenology to condensed matter systems. Some of the earliest accounts of the topic include Zamolodchikov's discovery of $E_8$ emergent behavior on the Ising model under longitudinal and transverse fields~\cite{Zamolodchikov:1987jf}, with subsequent experimental verification in cobalt niobate~\cite{E8_experiment}, about two and one decade ago, respectively. More recently, the interest extended from the $E_8$ to include other exceptional algebras such as $G_2$ and $F_4$, due to the potential impact these have in topological phases and, in particular, topological quantum computing~\cite{Yichen_G2,E8_quantum_Hall}.

Topological phases associated with the $G_2$ group, in particular, support low energy localized excitations whose behavior matches that of Fibonacci anyons, the simplest anyon capable of universal topological quantum computing~\cite{TQC_anyons}. Despite the interest, due to the inherent complexity of the group structure, proposals for realizing $G_2$-symmetric systems are still limited, and they are often too contrived for numerical or analytic calculations or experimental implementation.

The purpose of this paper is to explore $G_2$ physics from a model realization based on a standard magnetic structure. Working on 1D spin chains and following recent results by some of us~\cite{Li2022}, we embed a $G_2$-symmetric phase space within a $SU(2)$-symmetric magnetic model with local moments in the spin-3 representation. This idea brings the model closer to experimental relevance and also makes it amenable to efficient numerical analysis via $SU(2)$-symmetric non-Abelian density-matrix renormalization group (DMRG) methods.

Among the different characteristics of the $G_2$-symmetric phase space, we focus on those of a well-known \emph{integrable} point~\cite{OgievetskyWiegmann86,PokrovskiiTsvelik87,PokrovskyTsvelick89}. The typical association between isolated integrable models and critical points suggests that this point may be described by a conformal field theory (CFT) with $G_2$-symmetry. The simplest such theory would be a Wess--Zumino--Witten (WZW) $(G_2)_1$ CFT, whose single primary field satisfies the same fusion rules of Fibonacci anyons. The confirmation that this integrable model is described by a $(G_2)_1$ WZW CFT would also open the possibility for the existence of a critical phase with emergent $G_2$ symmetry, akin to the Uimin--Lai--Sutherland $SU(3)$-symmetric phase of the bilinear-biquadratic spin-1 $SU(2)$ chain~\cite{Uimin,Lai1974,Sutherland,Itoi1997}. Such systems would be prime candidates for coupled-wire constructions~\cite{sharmistha_coupled,Jeffrey_coupled_wire,Li2020} of 2D topological phases with Fibonacci anyons starting from an $SU(2)$-invariant system.

In principle, the integrable $G_2$ chain has been characterized in the 80's~\cite{OgievetskyWiegmann86,PokrovskiiTsvelik87,PokrovskyTsvelick89}. These classic references suggest that the low-energy physics of the model is indeed described by a $(G_2)_1$ WZW CFT. Yet, recent results have suggested that, albeit gapless, this model is not a CFT, but rather possesses two  low-energy excitation sectors with different spin-wave velocities~\cite{MartinsG2}. The fact that $G_2$ is not generated by a simply-laced algebra impacts the analysis of the analytic solution in a non-trivial way. The body of this work thus analyses the hypothesis that this integrable $G_2$ model is described by a $(G_2)_1$ WZW CFT via numerical methods. From exact diagonalization, we demonstrate that the low-energy spectrum of a small realization of the system does suggest the expected organization and degeneracy of a $(G_2)_1$ WZW CFT. Further characterization of the conformal data via non-Abelian DMRG, however, leads to mysteriously conflicting results. We estimate the central charge and primary field conformal dimensions from standard techniques such as energy spectrum analysis and entanglement entropy finite-size scaling, as well as a very recent new method using cyclic orbifolds and wavefunction overlaps~\cite{Shinsei_orbifold}, but find mismatching values. Reasons for the mismatches are put forward and include strong effects due to marginal perturbations and small system sizes, as well as the possibility that, indeed, the problem is not described by a CFT.

The paper is organized as follows. Section~\ref{sec:model} describes the embedding of $G_2$ chains in spin-3 $SU(2)$ symmetric ones, performs a cartographic analysis of what is known of the phase space of $G_2$ chains, and explores the finite-size spectrum of the integrable point via exact diagonalization. The case for a possible $(G_2)_1$ behavior is made. Section~\ref{sec:confdata} contains the bulk of our numerical characterization, including results from finite-size scaling of the energy spectrum, entanglement entropy, and wavefunction overlap. We conclude in Section~\ref{sec: conclusion}. We present pedagogical appendices on the conventions used to map $SU(2)$ and $G_2$ chains, a conformal field theory exploration of $(G_2)_1$, and its cyclic orbifolding. Finally, we present a reference appendix where our numerical analysis is redeployed on the Takhtajan--Babujian integrable point of spin-1 chains, whose continuum description in terms of an $SU(2)_2$ WZW CFT is well-established. The expected conformal data results are recovered in this case, demonstrating our numerical calculations are sound.

\section{$G_2$-invariance within spin-3 isotropic chains} \label{sec:model}

For our purposes, we write general $SU(2)$-invariant spin-3 chains in either of two forms, 
\begin{align}
    H&=\sum_{i=1}^{N}\sum_{n=0}^{6}\alpha_{n}\left(\mathbf{S}_{i}\cdot\mathbf{S}_{i+1}\right)^{n} \label{eq:spin3_SS}\\
    &=\sum_{i=1}^{N}\sum_{S=0}^{6}K_{S}P_{S}\left(\mathbf{S}_{i},\mathbf{S}_{i+1}\right), \label{eq:spin3_P}
\end{align}
where $\mathbf{S}_{i}=(S_i^a)$, $a=x,y,z$, are spin-3 operators acting at lattice site $i$, $\alpha_n$ and $K_S$ are constants, and $P_{S}\left(\mathbf{S}_{i},\mathbf{S}_{i+1}\right)$ are standard $SU(2)$ projectors on the multiplet $S$ of the sum of two local $\mathbf{S}_{i}$ and $\mathbf{S}_{i+1}$ spins; we use periodic boundary conditions overall. While Eq.~\eqref{eq:spin3_SS} displays the familiar form in terms of $SU(2)$-invariant bilinears and their powers, Eq.~\eqref{eq:spin3_P} is more convenient to analyze the symmetry structure of the parameter space.

This parameter space supports a $G_2$ embedding as follows~\cite{Li2022}: setting all $K_S=K$, the Hamiltonian is fine-tuned to a point of $SU(7)$ symmetry. Relaxing the constraint so that $K_1=K_3=K_5$, while $K_2=K_4=K_6$, combines the projectors into a $SO(7)$-symmetric system. Since  $SO(7) \supset G_2$, the symmetry can be further broken down. The Clebsch--Gordan series for two fundamental irreducible representations (irreps) of $G_2$ reads
\begin{equation}
\label{eq:tensorproduct77}
    \boldsymbol{7}\otimes \boldsymbol{7} = \boldsymbol{1} \oplus\boldsymbol{7} \oplus\boldsymbol{14} \oplus \boldsymbol{27}.
\end{equation}
Here we denote the irreps by their dimensions, for example with $\boldsymbol{7}$ being the smallest non-trivial irrep. To achieve these degeneracies from our spin-3 isotropic Hamiltonian, all it takes is to further loosen the constraints so that $K_3$ is not necessarily equal to $K_1=K_5$, that is, the constraints are $K_1=K_5$ and $K_2=K_4=K_6$.

Thus, if $\mathcal{P}_{\boldsymbol{\lambda}}$ are $G_2$ projectors on the space of the irreducible representation $\boldsymbol{\lambda}$, we have the identification
\begin{equation}
\begin{split}
\mathcal{P}_{\mathbf{1}}&=P_{S=0},\\
\mathcal{P}_{\mathbf{7}}&=P_{S=3},\\ 
\mathcal{P}_{\mathbf{14}}&=P_{S=1}+P_{S=5},\\ 
\mathcal{P}_{\mathbf{27}}&=P_{S=2}+P_{S=4}+P_{S=6},    
\end{split}
\end{equation}
leading to the general $G_2$-symmetric Hamiltonian
\begin{align}
    H_{G_2}&=\sum_{i=1}^{N}\sum_{\boldsymbol{\lambda}=\left\{ \mathbf{1},\mathbf{7},\mathbf{14},\mathbf{27}\right\} }F_{\boldsymbol{\lambda}}\mathcal{P}_{\boldsymbol{\lambda}}\left(\boldsymbol{\Lambda}_{i},\boldsymbol{\Lambda}_{i+1}\right) \label{eq:HG2}\\
    &= \sum_{i=1}^{N}\sum_{n=0}^{3}\beta_{n}\left(\boldsymbol{\Lambda}_{i}\cdot\boldsymbol{\Lambda}_{i+1}\right)^{n}. \label{eq:HG2_beta}
\end{align}
Here $\boldsymbol{\Lambda}_{i}=(\Lambda_i^a)$, $a=1,\dots,14$, are the $G_2$ generators acting on lattice site $i$, and $\boldsymbol{\Lambda}_{i}\cdot\boldsymbol{\Lambda}_{i+1}=\sum_a\Lambda_i^a\Lambda_{i+1}^a$. $F_{\boldsymbol{\lambda}}$ and $\beta_n$ control the parameter space of possible Hamiltonians; some explicit conversions are described in Appendix~\ref{app:G2_dictionary}. 
We choose the generators of $G_{2}$ to have the normalization $\text{Tr}\left((\Lambda^{a})^{\dagger}\Lambda^{b}\right)=2\delta_{ab}$.

The usefulness of this embedding is twofold: for one, it makes this family of $G_2$ Hamiltonians more physically attractive, and they can be realized in more conventional magnetism, albeit requiring high --- spin-3 --- spins. Furthermore, this embedding enables efficient numerical analysis based on well-developed methods that leverage the $SU(2)$ symmetry.

\begin{figure}[t]
    \centering
    \includegraphics[width=\columnwidth]{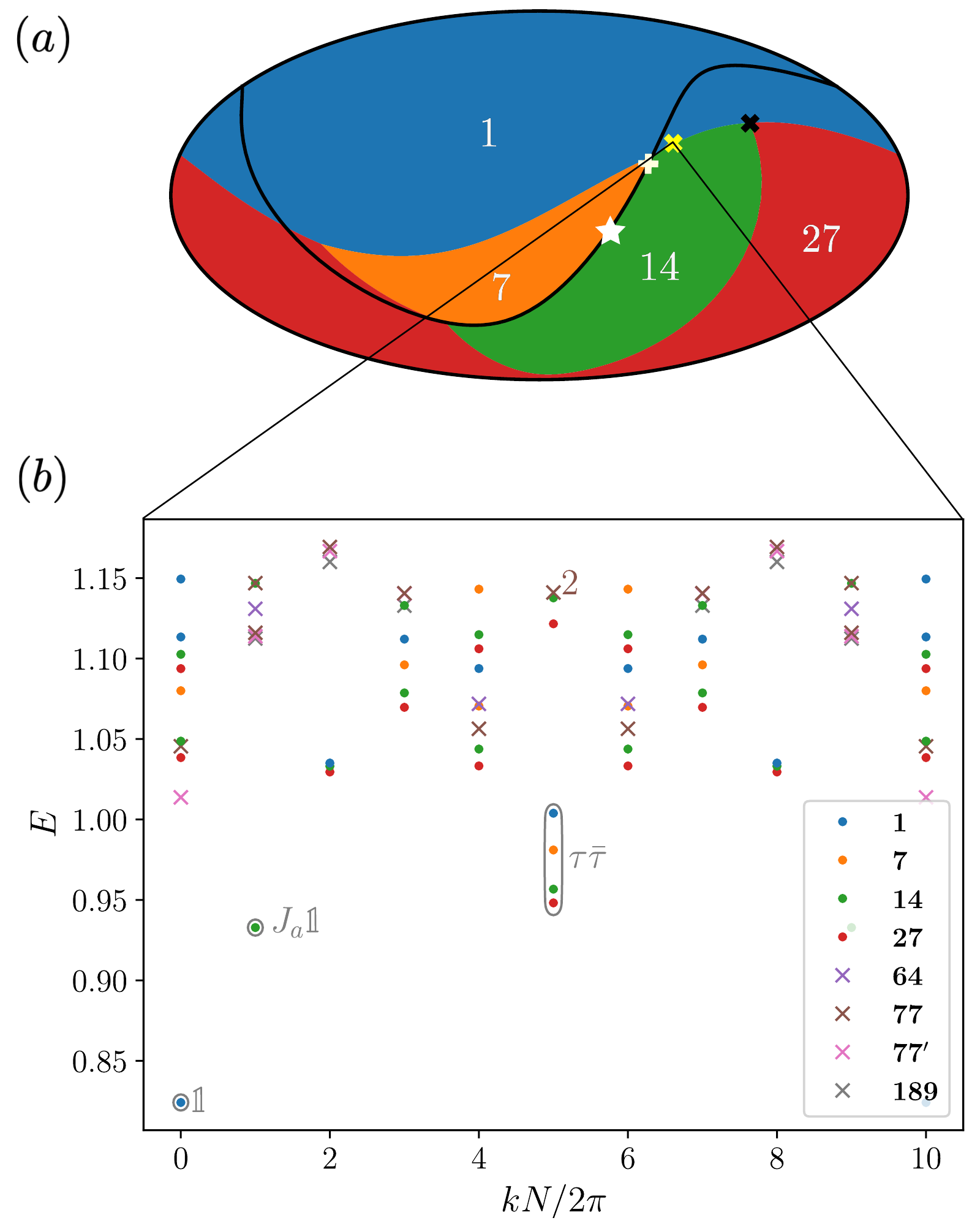}
    \caption{(a) Phase space of $G_2$-symmetric chains under a Mollweide projection. Numbers indicate the ground state representation for the 2-site problem. The black curve indicates the subspace of $SO(7)$-symmetric Hamiltonians, with the white star and cross marking the Reshetikhin $SO(7)_1$ and generalized ULS $SU(7)_1$ critical points - both integrable. The black cross marks the $G_2$-symmetric Hamiltonian found in Ref.~\cite{Li2022}, and the yellow cross indicates the $G_2$-symmetric integrable point of interest in this work. (b) Momentum-resolved exact-diagonalization spectrum for the integrable $G_2$ point with $N=10$ lattice sites. Irreducible representations are indicated in colors (the lowest ones with dots and higher ones with crosses). Multiplet collections and with corresponding hypothetical $(G_2)_1$ primary-field identifications are marked in grey. }
    \label{fig:G2_worldmap}
\end{figure}

From now, we stick only to the $G_2$ language, as it becomes the natural one for the problem. The phase space of Hamiltonian~\eqref{eq:HG2_beta} is 3-dimensional, as $\beta_0$ simply re-defines the ground-state energy. With a cartographic projection, the parameter space can be described in a 2-dimensional plane. Fig.~\ref{fig:G2_worldmap}{(a)} is a representation of the phase space using the Mollweide projection\footnote{The standard transformation from $\{\beta_i\}$ to the coordinates $(x,y)$ of the map reads $x=(2\sqrt2/\pi)\lambda\cos\theta$, $y=\sqrt2\sin\theta$, where $\lambda$ and $\theta$ are defined from $\beta_1=\beta\cos\phi\cos\lambda$, $\beta_2=\beta\cos\phi\sin\lambda$, $\beta_3=\beta\sin\phi$, $\beta=\sqrt{\beta_1^2+\beta_2^2+\beta_3^2}$, $2\theta+\sin2\theta=\pi\sin\phi$.}. The colors indicate the ground state of a 2-site problem. The space is mostly dominated by the blue singlet region, followed by the ``ferromagnetic'' red region (in the sense of the largest irrep having the lowest energy). The black line that cuts the parameter space corresponds to the line of $SO(7)$-symmetric Hamiltonians; the general many-body physics along this line is well-known and includes a ferromagnetic phase, an $SU(7)$-symmetric emergent phase, a gapped Haldane-like phase, and a dimerized phase~\cite{Tu2008}. The known critical points with conformal symmetry are indicated on the phase diagram, including the Reshetikhin $SO(7)_1$ point ($\bigstar$)~\cite{Reshetikhin1983,Reshetikhin1985}, and the generalized ULS $SU(7)_1$ point ($+$)~\cite{Uimin,Lai1974,Sutherland,Tu2008}. By studying phases with potential emergent symmetries, some of us previously explored a point in this phase space, labeled by the black cross ($\times$)~\cite{Li2022}.

The focus of analysis for this work corresponds to the point labeled by the \emph{yellow cross} in the parameter space shown in Fig.~\ref{fig:G2_worldmap}{(a)}. This point corresponds to an integrable model~\cite{OgievetskyWiegmann86,PokrovskiiTsvelik87,PokrovskyTsvelick89} we re-derive in Appendix~\ref{app:Integrability},
\begin{equation}
    H=\sum_{i=1}^{N} \left[\frac{1}{2}\boldsymbol{\Lambda}_{i}\cdot\boldsymbol{\Lambda}_{i+1}+\frac{5}{8}\left(\boldsymbol{\Lambda}_{i}\cdot\boldsymbol{\Lambda}_{i+1}\right)^{2}+\frac{75}{592}\left(\boldsymbol{\Lambda}_{i}\cdot\boldsymbol{\Lambda}_{i+1}\right)^{3}\right], \label{eq:G2_integrable}
\end{equation}
which we wrote in terms of powers of $G_2$-invariant bilinears. We note that even slightly changing the prefactors $1/2$, $5/8$, or $75/592$ in \eqref{eq:G2_integrable} immediately breaks the integrability of the model. 

We can also rewrite the model \eqref{eq:G2_integrable} in the more usual spin-3 language. Then the couplings are $\alpha_{0}=1122/925,\,\,\alpha_{1}=303/1850,\,\,\alpha_{2}=-3787/33300,\,\,\alpha_{3}=-712/24975,\,\,\alpha_{4}=-1/29970,\,\,\alpha_{5}=49/149850,\,\,\alpha_{6}=1/59940$, in terms of the Hamiltonian of Eq.~\eqref{eq:spin3_SS}, or $K_{0}=-4/37,\,\,K_{1}=K_{5}=0,\,\,K_{2}=K_{4}=K_{6}=24/37,\,\,K_{3}=18/37$, in terms of the projectors of Eq.~\eqref{eq:spin3_P}. In terms of $F$ [Eq.~\eqref{eq:HG2}], it is sufficient to use the identification that, generically, $F_{1}=K_{0},\,\,F_{7}=K_{3},\,\,F_{14}=K_{1}$ and $F_{27}=K_{2}$. Again, for a full dictionary on how to convert the couplings, we have tables in Appendix~\ref{app:G2_dictionary}.

Characterizing the physics of this model is important: besides the relevance for modeling or generating, Fibonacci anyon systems, as alluded to in the introduction, isolated integrable Hamiltonians are natural candidates to be critical points. The large continuous symmetry of the microscopic Hamiltonian Eq.~\eqref{eq:G2_integrable} suggests that, if this indeed corresponds to a critical point and displays conformal invariance, the odds are that the critical physics here is controlled by the $(G_2)_k$ class of WZW theories. If $k=1$, only one non-trivial primary field would be allowed, strongly limiting the allowed perturbations on this critical point. This suggests the existence of a larger region of the parameter space where an emergent $(G_2)_1$ liquid dominates the long-wavelength physics. This would happen in direct analogy to the $SU(3)_1$ low-energy behavior of spin-1 bilinear--biquadratic Hamiltonians close to the explicitly $SU(3)$-symmetric, and integrable, Uimin--Lai--Sutherland point~\cite{Uimin,Lai1974,Sutherland}.

For many years, results from the mathematical physics literature suggested that $(G_2)_1$ WZW CFT is the description for the low-energy behavior of integrable  model~\eqref{eq:G2_integrable}~\cite{PokrovskiiTsvelik87}. Yet, recently, a different analysis was put forward that suggests the description in terms of a product of two $c=1$ CFTs with different velocities~\cite{MartinsG2}. To explore numerically which results should be accurate, we consider the momentum-resolved spectrum via exact diagonalization of Hamiltonian~\eqref{eq:G2_integrable} for system size $N=10$ with periodic boundary conditions (which are used throughout this work). Fig.~\ref{fig:G2_worldmap}(b) displays the result; relying on the $G_2$ embedding in $SU(2)$, we use the total $S_z$ conservation to achieve larger system sizes and use branching rules~\cite{feger2014lieart} to reconstruct the sectors in terms of $G_2$ multiplets; symbols market as ``2'' mean there is a degeneracy between $G_2$ multiplets.

Some noteworthy observations follow: (i) the zero-momentum ground state is a singlet; (ii) the first excited state with finite momentum is a 14-dimensional multiplet; (iii) the lowest four multiplets at $\pi$ momentum form a 49-dimensional multiplet. While not a demonstration, these are suggestive of a $(G_2)_1$ CFT: as discussed in detail in Appendix~\ref{app:Orbi}, a holomorphic $(G_2)_1$ displays two conformal towers $\mathbbm{1}$ and $\tau$; at the lattice level, both holomorphic and anti-holomorphic parts are expected to mix as a tensor product. Comparing with the numerical results, both the ground-state singlet and the first finite-momentum 14-dimensional multiplet satisfy the expected degeneracies for an identity conformal tower and its first Kac--Moody descendant, generated by the $14$ current operators that generate $G_2$ at the CFT level (labeled $J_a$ in the figure). The 49-dimensional multiplet satisfies the expected degeneracy of the product of holomorphic and non-holomorphic $\tau \bar{\tau}$ conformal towers. Naturally, the lattice physics brings in marginal operators capable of lifting the correct degeneracy.\footnote{The same happens for the $SU(3)_1$ Uimin--Lai--Sutherland point for spin-1 chains~\cite{Itoi1997}. It is a known fact that including second-neighbor terms fine-tuned can counterbalance the effect of the marginal perturbation (see, e.g., the supplementary material of Ref.~\cite{Chen2015}. Yet, for the spin-1 problem, a single marginal operator is allowed, while three are possible for $G_2$. We did not succeed at tuning out the gaps in a system-size-independent manner.}

These observations serve as the starting point for our hypothesis that this integrable $G_2$ chain is a critical point described by a $(G_2)_1$ WZW CFT. For the rest of the paper, we analyze the scaling properties of states and observables to determine if our hypothesis can sustain itself.

\section{Conformal data characterization} \label{sec:confdata}

In this section, we focus on the characterization of key parameters of the integrable points based on the $(G_2)_1$ hypothesis. The key conformal parameters we need to determine from our model are the central charge $c$ and scaling dimension $\Delta_\tau$ of the only non-trivial primary operator $\tau$. For the sake of completeness, we state the expected values:
\begin{equation}
    c=\frac{14}{5}=2.8, \quad
    \Delta_\tau =\frac{4}{5}=0.8.
\end{equation}
For details on these and additional conformal data for the $(G_2)_1$ WZW CFT, see Appendix~\ref{app:Orbi}. 

\subsection{Energy scaling} \label{sec:energy}

\begin{figure*}[t]
    \centering
    \includegraphics[width=\textwidth]{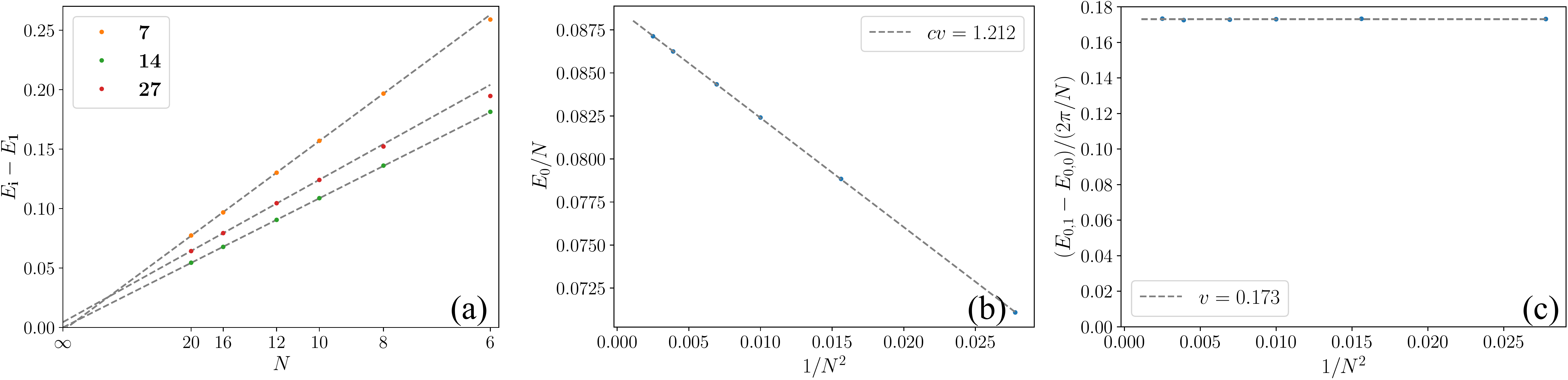}
    \caption{(a) Finite-size scaling of the lowest states in the 7, 14, and 27-dimensional multiplets; vanishing of the gaps is apparent but not conclusive. (b) Finite-size scaling of the ground-state energy, with factor $cv$ extracted  according to Eq.~\eqref{eq:E00}. (c) Extraction of velocity under the hypothesis that the 14-dimensional sector ground state corresponds to the first Kac descendant of the identity tower of $(G_2)_1$ (cf. $J_a\mathbb{I}$ in Fig.~\ref{fig:G2_worldmap})}
    \label{fig:energy_scalings}
\end{figure*}

The spectrum of Fig.~\ref{fig:G2_worldmap}(b) serves as an anchor for non-momentum-resolved but finer numerical analysis. Via $SU(2)$ non-Abelian DMRG, we extend our capacity for finite-size scaling up to $N=20$ (we use bond-dimension of $1000$ throughout, corresponding to $\sim 11000$ on a non-symmetry-preserving DMRG scheme) and analyze the standard CFT energy spectrum formulas:
\begin{align}
    &\frac{E_{0,0}\left(N\right)}{N}=\epsilon_{\infty}-\frac{\pi}{6N^{2}}cv \label{eq:E00},\\
    &\frac{E_{i,n}\left(N\right)-E_{0,0}\left(N\right)}{N}=\frac{2\pi v}{N^{2}}\left(\Delta_{i}+n\right). \label{eq:primaryenergy}
\end{align}
Here, $i$ indicates a conformal tower, and $n$ is an integer corresponding to descendant multiplets. Branching for high descendants can make the computation of $n$ unwieldy, but we only care about $n=1$ here. As for the other quantities, $\Delta_i$ is the (sum of holomorphic and anti-holomorphic) scaling dimension(s), $c$ is the central charge, and $v$ is a non-universal velocity. Further (logarithmic) corrections for the energies are known, but we do not consider them for the small system sizes we can reach. 

To access the CFT data, we first need to compute the non-universal spin-wave velocity $v$. To find it, we use the spectrum in Fig.~\ref{fig:G2_worldmap}(b) and proceed similarly to the approach of Ref.~\cite{Chen2015} as follows: under our hypothesis, the lowest-energy $\boldsymbol{14}$ multiplet corresponds to the first Kac-Moody descendent of the identity conformal tower $J_a\mathbbm{1}$, $a=1,...,14$ [cf. Fig.~\ref{fig:G2_worldmap}(b)]. We label this state with quantum numbers  $n,i=0,1$; its energy scaling, according to Eq.~\eqref{eq:primaryenergy}, reads
\begin{equation}
    \frac{E_{0,1}\left(N\right)-E_{0,0}\left(N\right)}{N}=\frac{2\pi v}{N^{2}}
\end{equation}
and gives us access to $v$. The green dots on Fig.~\ref{fig:energy_scalings}(a) shows our results for the energy finite-size scaling, and Fig.~\ref{fig:energy_scalings}(b) the extracted velocities. Typically, the velocity is fitted to $v(N)=v+a/N^2+b/N^4$~\cite{Chen2015}. We find here an unusual pattern, where~\footnote{We note that the difference of normalization between Martins~\cite{MartinsG2} and us is a factor of $37/12$, see Eq.~\eqref{eq:relationtoMartins}. If we multiply the extracted velocity by this factor we obtain $37v/12=0.53\approx\pi/6$, which is very close to the smaller velocity found by Martins.} $v(N)\approx0.1729(2)$ appears to change very little with system size $N$ for the system sizes we considered. Combining this velocity with the scaling of $E_{0,0}(N)$, from which we can extract $cv\approx 1.212(2)$ [cf. Fig~\ref{fig:energy_scalings}(c)], we obtain our first estimation of the central charge\footnote{Surprisingly, this matches the value of the central charge for $(G_2)_4$, as can be checked from Eq.~\eqref{eq:cc} in the Appendix. Still, the structure of states from exact diagonalization and the previous literature on Bethe ansatz solutions to this problem make $(G_2)_4$ a very unlikely scenario.} $c\approx7.01(1)$. This is exceedingly far from the expectation for $(G_2)_1$.

By targeting other $SU(2)$ sectors with our non-Abelian DMRG code, we can also study the finite-size scaling of further relevant states. The orange and red  dots on Fig~\ref{fig:energy_scalings}(a) includes states in the multiplets $\textbf{7}$ and $\textbf{27}$. These states are supposed to merge, as $N\rightarrow \infty$, in a single $49$-dimensional multiplet of $\tau \times \bar{\tau}$. Without canceling the contributions from marginal operators, we can obtain a range of possible values for the scaling dimension between $\Delta_\tau\sim 1.1$ and $\Delta_\tau\sim 1.6$. Given the off behavior for $c$ above, it is unsurprising that we also find an off estimate for the scaling dimension. 

Generally, the results of this section are puzzling. The conformal data is way off the expectation for that of $(G_2)_1$. 

These numbers are strongly dependent on our capacity to estimate the velocity $v$, and we decidedly observe an unusually flat behavior for how $v$ scales with system size. Yet, the method presented here is known to obtain trustworthy results in other situations. We successfully tested it on the $SU(2)_2$ Takhtajan--Babujian (TB) point of spin-1 bilinear--biquadratic chains (cf. Appendix~\ref{app:TKTB}) to extract the central charge $c_\mathrm{TB}=3/2$~\cite{francesco2012conformal}, for example. It has also been previously used to analyze the $SU(3)_1$ behavior in spin-1 bilinear--biquadratic chains and, under somewhat less certain conditions, in spin-2 chains~\cite{Chen2015,Li2022}. 

In what follows, we cross-check these results by determining conformal parameters via two other fully independent methodologies.

\subsection{Entanglement entropy}
Another well-known method of obtaining the central charge of a CFT involves the finite-size behavior of the entanglement spectrum~\cite{CalabreseCardy_EE}
\begin{equation}
    S\left(j,N\right)=\frac{c}{3}\ln\left[\frac{N}{\pi}\sin\left(\frac{\pi j}{N}\right)\right]+S_{0}, \label{eq:EE}
\end{equation}
where $S_{0}$ is a constant asymptotic and non-universal contribution, and $j$ is a size of bipartition of the lattice. While this method has the advantage of bypassing the need to compute the velocity $v$, the DMRG convergence can make this method unreliable when the problem of interest contains phases close by with competing ground states~\cite{Chen2012,Chen2015}.

\begin{figure}[t]
    \centering
    \includegraphics[width=\columnwidth]{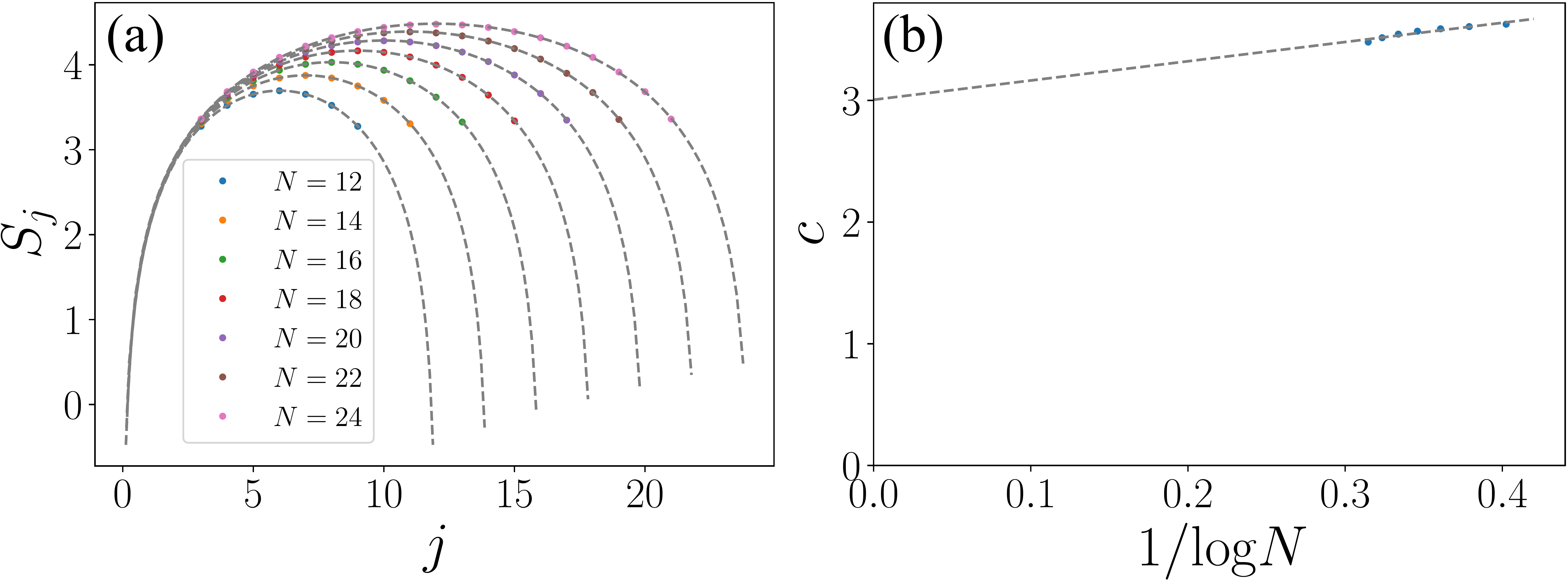}
    \caption{(a) Finite-size scaling of entanglement entropy and fitting according to Eq.~\eqref{eq:EE}. (b) Finite-size scaling of the extracted values of central charge $c$.} 
    \label{fig:EE}
\end{figure}

The results of our calculations are shown in Fig.~\ref{fig:EE}(a) and (b). Despite good fits to the expected functional shape of the entanglement entropy, the central charge of the model shows a strong dependence on system size. In the absence of a systematic model or theory for this, we studied different fitting functions, both linear and non-linear, obtaining values for the extrapolated central charge varying from $c\sim2.8$ to $c\sim3.4$. These are much closer to the expected $c=2.8$ for $(G_2)_1$ than the results from the previous section. 

The results from this method are strikingly discrepant from the energy finite-size scaling of the previous session. The only known $G_2$-symmetric CFT with a central charge below $3.4$ is indeed the level-1 CFT of our hypothesis. Yet, it is hard to say if the difficulty in fitting we observe is only due to finite-size effects or if it comes to be due to the system not being well described by a pure CFT. Further analysis is necessary to determine the nature of the system conclusively.

\subsection{Wavefunction overlap}

A new method has recently been introduced to extract the CFT data from numerical realizations of the critical theory~\cite{Shinsei_orbifold}. The process involves considering two identical periodic copies of the system of interest at a size $N$, described by a CFT with (\emph{non-chiral}) primary fields $\phi_{\alpha}^{1}$ and $\phi_{\beta}^{2}$, and a third periodic copy of size $2N$ and primary fields $\phi_{\gamma}^{3}$. Then, one considers the overlap 
\begin{equation}
    A_{\alpha\beta\gamma}\equiv\left\langle \phi_{\gamma}^{3}|\phi_{\alpha}^{1}\phi_{\beta}^{2}\right\rangle. \label{eq:overlaps}
\end{equation}
Leveraging a process of cyclic orbifolding the original CFT of interest, it is possible to show that finite-size realizations of the CFT enforce scaling laws on the overlaps above that are fixed by the conformal data of the CFT. Details are discussed in the original work~\cite{Shinsei_orbifold} and summarized in Appendix~\ref{app:Orbi}. The simplest overlap to consider involves the identity primary state,
\begin{align}
    A_{\mathbbm{1}\mathbbm{1}\mathbbm{1}}&\propto N^{-c/8}+.... \label{eq:A111}
\end{align}
Here $\mathbbm{1}$ indicates the identity conformal tower, and the ellipses correspond to sub-dominant contributions from descendants. This overlap offers a direct venue to the central charge $c$ by finite-size scaling. The correspondence between the CFT and lattice states is very simple in this case, and one simply follows

\begin{equation}
    \left\langle \mathbbm{1}^{3}|\mathbbm{1}^{1}\mathbbm{1}^{2}\right\rangle \leftrightarrow\left\langle 3:S=0,0|1:S=0,0;2:S=0,0\right\rangle, 
\end{equation}
where the notation $\left|n:S,i\right\rangle $ indicates the $i$-th state in of total angular momentum $S$ for spin-chain $n$ (with lengths $N$ for $n=1,\,2$ and $2N$ for $n=3$). $i=0$ indicates the ground state. In other words, we identify $\left|\mathbbm{1}^n\right\rangle \leftrightarrow\left|n:S=0,0\right\rangle $.

\begin{figure}[t]
    \centering
    \includegraphics[width=\columnwidth]{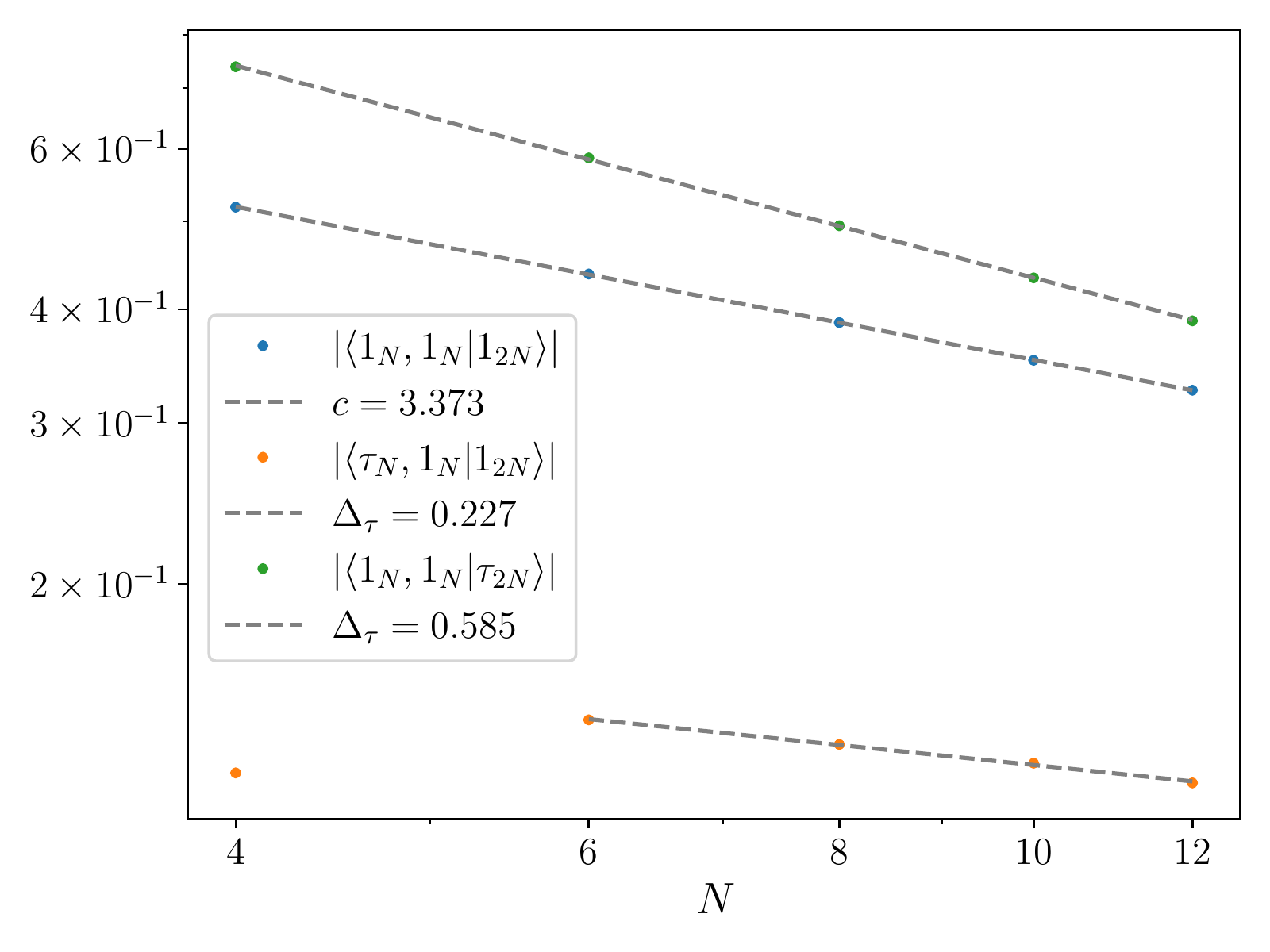}
    \caption{(blue) Wavefunction overlap extracting the central charge according to ground state overlap as in Eq.~\eqref{eq:A111}. (green) and (orange) similarly correspond to overlaps extracting the Fibonacci primary scaling dimension according to Eqs.~\eqref{eq:A11gamma} and \eqref{eq:Agamma11}, respectively. States $\tau$ used for the overlap computation corresponds to the singlet state of the multiplet collection indicated by $\tau \bar{\tau}$ in grey in Fig.~\ref{fig:G2_worldmap}.} 
    \label{fig:overlaps}
\end{figure}

Fig.~\ref{fig:overlaps}(blue) displays the results of our numerical analysis. The all-identity conformal tower overlap estimates a central charge $c\approx 3.37(1)$, closer to the expected value of $(G_2)_1$ than to the unexpected value found in Sec.~\ref{sec:energy}, but not satisfactorily close to $2.8$, our original expectation. Given the small system sizes we can reach, the corrections in the ellipses might be important and may the responsible for the discrepancy with the expected result. To show that our method is nevertheless trustworthy, we repeat the analysis in Appendix \ref{app:TKTB}.

Interestingly, the method of wave-function overlaps also allows for a simple way to estimate scaling dimensions of primary operators, provided one can accurately access excited states numerically. Two noteworthy results are
\begin{align}
    \frac{A_{\mathbbm{1}\mathbbm{1}\gamma}}{A_{\mathbbm{1}\mathbbm{1}\mathbbm{1}}}&\propto N^{-\Delta_{\gamma}/2}+..., \label{eq:A11gamma} \\ 
    \frac{A_{\gamma\mathbbm{1}\mathbbm{1}}}{A_{\mathbbm{1}\mathbbm{1}\mathbbm{1}}}&\propto N^{-\Delta_{\gamma}/2}\left(1+\frac{a}{N}+...\right). \label{eq:Agamma11}
\end{align}
 On the last ratio above, $a$ is a constant, and the first sub-dominant power law is controlled by $(\gamma,\hat{1})$, a primary field in the twist sector of the cyclic orbifold of the original CFT. These overlaps require the calculation of an excited state (corresponding to the primary $\gamma$) on chain copy $n=1,2$, of size $N$, or $n=3$ of size $2N$. 

In practice, computing Eq.~\eqref{eq:Agamma11} from lattice numerics brings some extra subtlety. While $G_2$ degeneracy is explicit at the lattice level, Fig.~\ref{fig:G2_worldmap}(b) suggests splittings of the expected conformal towers due to marginal perturbations. One is forced to pick a given multiplet by hand when looking for states corresponding, say, to the $\tau$ conformal tower. Furthermore, since our spin chain and DMRG routine rely on $SU(2)$ symmetry to enhance computational capacity, our  diagonalization targets sectors of fixed total angular momentum. As states of different total angular momentum are orthogonal by construction, one is forced to consider overlaps always within the same multiplet. For example, for $A_{\mathbbm{1}\mathbbm{1}\gamma}$, one identifies
\begin{equation}
    \left\langle \mathbbm{1}^{3}|\mathbbm{1}^{1}\tau^{2}\right\rangle \leftrightarrow\left\langle 3:S=0,0|1:S=0,0;2:S=0,1\right\rangle. 
\end{equation}
Here, we see that $\left|\tau^2\right\rangle \leftrightarrow\left|2:S=0,1\right\rangle$, i.e., the first excited state ($i=1$) for the $S=0$ sector on the lattice, matches the first blue dot at momentum $\pi$ in Fig.~\ref{fig:G2_worldmap}(b). Due to small system sizes, the difficulty of convergence of DMRG for excited states, and the issue with marginal perturbations, the overlap involving $\tau$ states are much less trustworthy, and further discrepancies are then expected when trying to estimate $\Delta_\tau$. 

The overlaps $A_{\mathbbm{1}\mathbbm{1}\tau}/A_{\mathbbm{1}\mathbbm{1}\mathbbm{1}}$ and $A_{\tau\mathbbm{1}\mathbbm{1}}/A_{\mathbbm{1}\mathbbm{1}\mathbbm{1}}$ are shown inFig.~\ref{fig:overlaps} green and orange, respectively. The corresponding estimations for scaling dimensions return $\Delta_\tau \approx 0.227$ and $\Delta_\tau \approx 0.585$. Indeed, these results are far-off the expected values and distinct from each other as well.

\section{Discussion and conclusion} \label{sec: conclusion}

We report on a numerical analysis of an integrable $G_2$-symmetric 1D chain. Embedding this model in the space of $SU(2)$-symmetric spin-3 systems, we bring the $G_2$ model closer to physical relevance, as well as enable efficient numerical analysis via non-Abelian density matrix renormalization group. 

Both simplicity and a momentum-resolved spectrum from exact diagonalization suggest that the integrable system may be described by a $(G2)_1$ WZW CFT. Yet, when different methods are used to extract the conformal information from the system (central charge and conformal dimension), very discrepant results are obtained. The methods explored here are all based on finite-size scaling of different quantities, namely of the energy spectrum of the chain, of its entanglement entropy, and of the wavefunction overlap of states that should match primary fields in the CFT. The last method is a very recently introduced approach to extracting conformal data from numerical studies.

Since the discrepant results we find are very unexpected, we test our methodologies against a different integrable system whose conformal field theory description is well established. In Appendix~\ref{app:TKTB}, we deploy the very same methods for the Takhtajan--Babujian  point of spin-1 bilinear--biquadratic spin chains, known to correspond to an $SU(2)_2$ WZW CFT~\cite{Takhtajan1982,Babujian_PhysLetA_1982,Affleck_TKTB}. All of our numerical data point to the same central charge $c\approx 3/2$, the expected value, suggesting that our methods are implemented in a reliable way. Still, scaling dimensions are not so easily obtained numerically for this system~\cite{Affleck_TKTB}, and the fact that the lowest state at momentum $\pi$ lies in the sector of $S=1$ makes the wavefunction overlap method hard to deploy. Given the verified trustworthiness of the deployed methodologies, we are left with a mystery at hand. 

We may leave a few final observations and hypotheses to explain our findings. First, the energy-scaling analysis depends on an estimation of the spin-wave velocity, and the latter shows an unexpectedly flat behavior in this system. A second point is that the entanglement entropy and wavefunction overlap both estimate the central charge closer to $2.8$, perhaps indicating that the system sizes we are able to probe are not large enough to display the WZW behavior. The existence of marginal operators can also be responsible for tainting our numerical results. A final possibility is that all numerical analyses are actually sound, and this integrable system is, in fact, not a CFT. While this is may be unexpected (even more for a point in isolation in phase space), this could potentially explain why different methodologies to analyze conformal data return mismatching results, and be in line with the analysis performed recently by Martins~\cite{MartinsG2}. Further studies are needed to shed light on the real physical nature of this $G_2$ integrable system and its vicinity in phase space.

\acknowledgements
We thank Frank G\"ohmann, Hosho Katsura, Shinsei Ryu, and Rodrigo Pereira for useful discussions. The project is supported by China Postdoctoral Science Foundation (Grant No. 2022M711868). C. L. is also supported by the International Postdoctoral Exchange Fellowship Program and the Shuimu Tsinghua Scholar Program. 

\appendix
\section{$G_2$ dictionaries}\label{app:G2_dictionary}
In this Appendix, we write down the mapping of the spin-3 model to the $G_{2}$ invariant language (and the inverse mappings as well). It is enough, for this purpose, to consider a two-site problem. 

\subsection{Coupling Dictionary}
We first determine the constraints on the couplings $\alpha_{n}$ such that the Hamiltonian is $G_{2}$ symmetric (see the Hamiltonian in terms of  $\alpha_{n}$ in Eq.~\ref{eq:spin3_SS}). Let us keep $\alpha_{0},\alpha_{1},\alpha_{2}$ and $\alpha_{3}$ as free parameters. The couplings $\alpha_{4},\alpha_{5}$ and $\alpha_{6}$ read

\begin{align}
\alpha_{4} & =\frac{35350}{65144169}\alpha_{1}-\frac{198}{9929}\alpha_{2}+\frac{67426}{804249}\alpha_{3},\\
\alpha_{5} & =-\frac{1}{6561}\alpha_{1}-\frac{1}{81}\alpha_{3},\\
\alpha_{6} & =-\frac{680}{65144169}\alpha_{1}+\frac{1}{9929}\alpha_{2}-\frac{842}{804249}\alpha_{3}.
\end{align}
The constraints for the couplings $K_{S}$ (Eq.~\ref{eq:spin3_P}) to have a $G_{2}$-symmetric Hamiltonian are

\begin{equation}
K_{5}=K_{1},\,\,K_{6}=K_{4}=K_{2}.
\end{equation}
We are left with $K_{0},\,K_1,\,K_{2}$, and $K_{3}$ as arbitrary parameters.

We list the conversion from the different couplings constants.  We write down the couplings in terms of $\beta$ in Table~\ref{tab:Conversion-beta} (see Eq. ~\ref{eq:HG2_beta}),
of $F$ in Table~\ref{tab:Conversion-F} (Eq.~\ref{eq:HG2}), of $K$ (Eq.~\ref{eq:spin3_P}) in Table~\ref{tab:Conversion-K} and, finally, in terms of $\alpha$ (Eq.~\ref{eq:spin3_SS}) in Table~\ref{tab:Conversion-alpha}.

\begin{center}
\begin{table}
\begin{centering}
\begin{tabular}{|c|c|}
\hline 
 & $\beta$\tabularnewline
\hline 
$K$ & $K_{0}=\beta_{0}-4\beta_{1}+16\beta_{2}-64\beta_{3}$\tabularnewline
 & $K_{1}=\beta_{0}$\tabularnewline
 & $K_{2}=\beta_{0}+\frac{2}{3}\beta_{1}+\frac{4}{9}\beta_{2}+\frac{8}{27}\beta_{3}$\tabularnewline
 & $K_{3}=\beta_{0}-2\beta_{1}+4\beta_{2}-8\beta_{3}$\tabularnewline
\hline 
$\alpha$ & $\alpha_{0}=\beta_{0}+\frac{176}{75}\beta_{1}+\frac{1144}{225}\beta_{1}-\frac{3344}{135}\beta_{3}$\tabularnewline
 & $\alpha_{1}=\frac{9}{25}\beta_{1}+\frac{58}{25}\beta_{2}-\frac{868}{75}\beta_{3}$\tabularnewline
 & $\alpha_{2}=-\frac{433}{1350}\beta_{1}-\frac{989}{2025}\beta_{1}+\frac{16874}{6075}\beta_{3}$\tabularnewline
 & $\alpha_{3}=-\frac{31}{900}\beta_{1}-\frac{1133}{4050}\beta_{2}+\frac{7843}{6075}\beta_{3}$\tabularnewline
\hline 
$F$ & $F_{1}=\beta_{0}-4\beta_{1}+16\beta_{2}-64\beta_{3}$\tabularnewline
 & $F_{7}=\beta_{0}-2\beta_{1}+4\beta_{2}-8\beta_{3}$\tabularnewline
 & $F_{14}=\beta_{0}$\tabularnewline
 & $F_{27}=\beta_{0}+\frac{2}{3}\beta_{1}+\frac{4}{9}\beta_{2}+\frac{8}{27}\beta_{3}$\tabularnewline
\hline 
\end{tabular}
\par\end{centering}
\caption{Conversion of the couplings, written in terms of $\beta$.~\label{tab:Conversion-beta}}
\end{table}
\par\end{center}

\begin{center}
\begin{table}
\begin{centering}
\begin{tabular}{|c|c|}
\hline 
 & $F$\tabularnewline
\hline 
$K$ & $K_{0}=F_{1}$\tabularnewline
 & $K_{1}=F_{14}$\tabularnewline
 & $K_{2}=F_{27}$\tabularnewline
 & $K_{3}=F_{7}$\tabularnewline
\hline 
$\alpha$ & $\alpha_{0}=\frac{99F_{1}}{175}-\frac{6F_{14}}{5}+\frac{517F_{27}}{175}-\frac{33F_{7}}{25}$\tabularnewline
 & $\alpha_{1}=\frac{6F_{1}}{25}-\frac{21F_{14}}{50}+\frac{63F_{27}}{100}-\frac{9F_{7}}{20}$\tabularnewline
 & $\alpha_{2}=-\frac{431F_{1}}{6300}+\frac{281F_{14}}{1350}-\frac{4129F_{27}}{12600}+\frac{203F_{7}}{1080}$\tabularnewline
 & $\alpha_{3}=-\frac{122F_{1}}{4725}+\frac{343F_{14}}{5400}-\frac{1007F_{27}}{12600}+\frac{19F_{7}}{450}$\tabularnewline
\hline 
$\beta$ & $\beta_{0}=F_{14}$\tabularnewline
 & $\beta_{1}=\frac{1}{28}F_{1}-\frac{1}{4}F_{7}-\frac{3}{4}F_{14}+\frac{27}{28}F_{27}$\tabularnewline
 & $\beta_{2}=-\frac{F_{1}}{28}+\frac{5}{16}F_{7}-F_{14}+\frac{81}{112}F_{27},$\tabularnewline
 & $\beta_{3}=-\frac{3}{112}F_{1}+\frac{3}{32}F_{7}-\frac{3}{16}F_{14}+\frac{27}{224}F_{27}$\tabularnewline
\hline 
\end{tabular}
\par\end{centering}
\caption{Conversion of the couplings, written in terms of $F$.~\label{tab:Conversion-F}}
\end{table}
\par\end{center}

\begin{center}
\begin{table}
\begin{centering}
\begin{tabular}{|c|c|}
\hline 
 & $K$\tabularnewline
\hline 
$\alpha$ & $\alpha_{0}=\frac{99K_{0}}{175}-\frac{6K_{1}}{5}+\frac{517K_{2}}{175}-\frac{33K_{3}}{25}$\tabularnewline
 & $\alpha_{1}=\frac{6K_{0}}{25}-\frac{21K_{1}}{50}+\frac{63K_{2}}{100}-\frac{9K_{3}}{20}$\tabularnewline
 & $\alpha_{2}=-\frac{431K_{0}}{6300}+\frac{281K_{1}}{1350}-\frac{4129K_{2}}{12600}+\frac{203K_{3}}{1080}$\tabularnewline
 & $\alpha_{3}=-\frac{122K_{0}}{4725}+\frac{343K_{1}}{5400}-\frac{1007K_{2}}{12600}+\frac{19K_{3}}{450}$\tabularnewline
\hline 
$F$ & $F_{1}=K_{0}$\tabularnewline
 & $F_{7}=K_{3}$\tabularnewline
 & $F_{14}=K_{1}$\tabularnewline
 & $F_{27}=K_{2}$\tabularnewline
\hline 
$\beta$ & $\beta_{0}=K_{1}$\tabularnewline
 & $\beta_{1}=\frac{1}{28}K_{0}-\frac{3}{4}K_{1}+\frac{27}{28}K_{2}-\frac{1}{4}K_{3}$\tabularnewline
 & $\beta_{2}=-\frac{1}{28}K_{0}-K_{1}+\frac{81}{112}K_{2}+\frac{5}{16}K_{3}$\tabularnewline
 & $\beta_{3}=-\frac{3}{112}K_{0}-\frac{3}{16}K_{1}+\frac{27}{224}K_{2}+\frac{3}{32}K_{3}$\tabularnewline
\hline 
\end{tabular}
\par\end{centering}
\caption{Conversion of the couplings, written in terms of $K$.~\label{tab:Conversion-K}}
\end{table}
\par\end{center}

\begin{center}
\begin{table}
\begin{centering}
\begin{tabular}{|c|c|}
\hline 
 & $\alpha$\tabularnewline
\hline 
\hline 
$K$ & $K_{1}=\alpha_{0}+\frac{2412190}{804249}\alpha_{1}+\frac{74052}{9929}\alpha_{2}+\frac{298144}{9929}\alpha_{3}$\tabularnewline
 & $K_{0}=\alpha_{0}+\frac{4832980}{804249}\alpha_{1}+\frac{310032}{9929}\alpha_{2}-\frac{433856}{9929}\alpha_{3}$\tabularnewline
 & $K_{2}=\alpha_{0}-\frac{19730}{9929}\alpha_{1}+\frac{36612}{9929}\alpha_{2}-\frac{62856}{9929}\alpha_{3}$\tabularnewline
 & $K_{3}=\alpha_{0}-\frac{3698390}{804249}\alpha_{1}+\frac{147492}{9929}\alpha_{2}-\frac{597656}{9929}\alpha_{3}$\tabularnewline
\hline 
$F$ & $F_{1}=\alpha_{0}+\frac{4832980\text{\ensuremath{\alpha_{1}}}}{804249}+\frac{310032\text{\ensuremath{\alpha_{2}}}}{9929}-\frac{433856\alpha_{3}}{9929}$\tabularnewline
 & $F_{7}=\alpha_{0}-\frac{3698390\alpha_{1}}{804249}+\frac{147492\text{\ensuremath{\alpha_{2}}}}{9929}-\frac{597656\alpha_{3}}{9929}$\tabularnewline
 & $F_{14}=\alpha_{0}+\frac{2412190\text{\ensuremath{\alpha_{1}}}}{804249}+\frac{74052\alpha_{2}}{9929}+\frac{298144\alpha_{3}}{9929}$\tabularnewline
 & $F_{27}=\alpha_{0}-\frac{19730\alpha_{1}}{9929}+\frac{36612\alpha_{2}}{9929}-\frac{62856\alpha_{3}}{9929}$\tabularnewline
\hline 
$\beta$ & $\beta_{0}=\text{\ensuremath{\alpha_{0}}}+\frac{2412190}{804249}\alpha_{1}+\frac{74052}{9929}\alpha_{2}+\frac{298144}{9929}\alpha_{3}$\tabularnewline
 & $\beta_{1}=-\frac{500665}{178722}\alpha_{1}-\frac{46035}{9929}\alpha_{2}-\frac{150300}{9929}\alpha_{3}$\tabularnewline
 & $\beta_{2}=-\frac{6528445}{1072332}\alpha_{1}-\frac{12555}{9929}\alpha_{2}-\frac{514875}{9929}\alpha_{3}$\tabularnewline
 & $\beta_{3}=-\frac{498265}{357444}\alpha_{1}-\frac{15795}{39716}\alpha_{2}-\frac{215775}{19858}\alpha_{3}$\tabularnewline
\hline 
\end{tabular}
\par\end{centering}
\caption{Conversion of the couplings, written in terms of $\alpha$.~\label{tab:Conversion-alpha}}
\end{table}
\par\end{center}

\subsection{$G_{2}$ projectors}

To link the different representations of the model, we also write down the projectors $\mathcal{P}_{\mathbf{n}}$ in terms of the Casimir
$\mathcal{C}_{2}\equiv\left(\boldsymbol{\mathbf{\Lambda}}_{i}+\boldsymbol{\mathbf{\Lambda}}_{i+1}\right)^{2}$ to the projectors of the Clebsch--Gordan series of two $G_2$ fundamental irreps. They read
\begin{align}\label{eq:projectors}
    \mathcal{P}_{\mathbf{1}}={}&-\frac{3}{896}\left[\mathcal{C}_{2}-4\right]\left[\mathcal{C}_{2}-8\right]\left[\mathcal{C}_{2}-\frac{28}{3}\right] \nonumber \\
    \mathcal{P}_{\mathbf{7}}={}&\frac{3}{256}\mathcal{C}_{2}\left[\mathcal{C}_{2}-8\right]\left[\mathcal{C}_{2}-\frac{28}{3}\right] \nonumber \\
    \mathcal{P}_{\mathbf{14}}={}&-\frac{3}{128}\mathcal{C}_{2}\left[\mathcal{C}_{2}-4\right]\left[\mathcal{C}_{2}-\frac{28}{3}\right] \nonumber \\
    \mathcal{P}_{\mathbf{27}}={}&\frac{27}{1792}\mathcal{C}_{2}\left[\mathcal{C}_{2}-4\right]\left[\mathcal{C}_{2}-8\right].
\end{align}

Recall that we normalize the generators of $G_{2}$ according to $\text{Tr}\left(\Lambda_{\alpha}^{\dagger}\Lambda_{\beta}\right)=2\delta_{\alpha\beta}$. For irreps ${\textbf{1},\,\textbf{7},\,\textbf{14},\,\textbf{27}}$, we have, respectively $\mathcal{C}_{2}=\left\{ 0,4,8,\frac{28}{3}\right\}$. 

\section{$G_{2}$ Integrable point}\label{app:Integrability}

In this appendix we derive the integrable $G_2$ spin chain \eqref{eq:G2_integrable}. We do so by first using the general construction of a rational R-matrix given by MacKay~\cite{MacKay91}, which then serves as starting point in the quantum inverse scattering method~\cite{KorepinBogoliubovIzergin93,SamajBajnok13}. We consider a chain with fundamental $G_2$-representations at each lattice site. The tensor product on two neighbouring sites can be decomposed as given in \eqref{eq:tensorproduct77}, with the corresponding eigenvalues of the quadratic Casimir operator given in the previous section. Following the general construction of rational R-matrices in irreducible representations~\cite{MacKay91} we directly obtain
\begin{equation}
\begin{split}
R(\lambda)={}&\mathcal{P}_{\mathbf{1}}+\frac{\lambda+\eta}{\lambda-\eta}\frac{\lambda+\tfrac{\eta}{6}}{\lambda-\tfrac{\eta}{6}}\frac{\lambda-\tfrac{2\eta}{3}}{\lambda+\tfrac{2\eta}{3}}\mathcal{P}_{\mathbf{7}}\\
{}&+\frac{\lambda+\eta}{\lambda-\eta}\mathcal{P}_{\mathbf{14}}+\frac{\lambda+\eta}{\lambda-\eta}\frac{\lambda+\tfrac{\eta}{6}}{\lambda-\tfrac{\eta}{6}}\mathcal{P}_{\mathbf{27}},
\end{split}
\label{eq:Rmatrix}
\end{equation}
where the projectors are explicitly given in terms of the Casimir operator in \eqref{eq:projectors}, $\lambda$ denotes the rapidity, and $\eta$ is a free parameter. It is straightforward to show that the R-matrix satisfies the Yang--Baxter equation 
\begin{equation}
    R_{12}(\lambda-\mu)R_{13}(\lambda)R_{23}(\mu)=R_{23}(\mu)R_{13}(\lambda)R_{12}(\lambda-\mu),
\end{equation}
where the subindex denotes on which of the factors in the tensor product $\mathbf{7}\otimes\mathbf{7}\otimes\mathbf{7}$ the R-matrix acts non-trivially. The R-matrix further satisfies the normalisation 
\begin{equation}
    R(\lambda=0)=\mathcal{P}_{\mathbf{1}}-\mathcal{P}_{\mathbf{7}}-\mathcal{P}_{\mathbf{14}}+\mathcal{P}_{\mathbf{27}}=P,
\end{equation}
with the permutation operator $P$, and $R(\lambda)\big|_{\eta=0}=1$. We note that the R-matrix \eqref{eq:Rmatrix} has been obtained previously by Ogievetsky~\cite{Ogievetsky86} and its q-deformation by Kuniba~\cite{Kuniba90}. 

The R-matrix \eqref{eq:Rmatrix} can now be used as input to construct an integrable  Hamiltonian. We follow Ref.~\onlinecite{SamajBajnok13}, noting that the relation to our convention is provided by $S_{ab}^{cd}\leftrightarrow R_{ab}^{cd}$. We define the transfer matrix via
\begin{equation}
\tau_{\sigma_1\ldots\sigma_N}^{\sigma_1'\ldots\sigma_N'}(\lambda)=
R_{\sigma_1\gamma_2}^{\sigma_1'\gamma_1}(\lambda)R_{\sigma_2\gamma_3}^{\sigma_2'\gamma_2}(\lambda)\cdots R_{\sigma_N\gamma_1}^{\sigma_N'\gamma_N}(\lambda)
\end{equation}
acting on the Hilbert space $\mathcal{H}=\bigotimes_{i=1}^N V$ with $V=\mathbf{7}$. Now using $R_{\alpha\beta}^{\gamma\delta}(\lambda=0)=P_{\alpha\beta}^{\gamma\delta}=\delta_{\alpha}^{\delta}\delta_\beta^\gamma$ we get 
\begin{equation}
H=\frac{\partial}{\partial\lambda}\ln[\tau(\lambda)]\Big|_{\lambda=0}=\sum_{i=1}^N H_i,\quad
H_i=\frac{\partial}{\partial\lambda}R(\lambda)\Big|_{\lambda=0}\,P,
\label{eq:QISMhamiltonian}
\end{equation}
where the factor $P$ ensures the correct indices as compared to  Ref.~\onlinecite{SamajBajnok13}. Periodic boundary conditions are imposed. Using \eqref{eq:Rmatrix} in \eqref{eq:QISMhamiltonian} as well as  $\mathcal{C}_2=4+2\mathbf{\Lambda}_i\cdot\mathbf{\Lambda}_{i+1}$ and setting  $\eta=37/2$ we arrive at Eq.~\eqref{eq:G2_integrable}, up to an additive constant. This construction, albeit the Hamiltonian was less explicitly stated, has also been demonstrated in~\cite{OgievetskyWiegmann86,PokrovskiiTsvelik87,PokrovskyTsvelick89}. The link to the conventions used by Martins is, up to an additive constant, provided by (see Eq.~(20) in Ref.~\onlinecite{MartinsG2})
\begin{equation}
    H_\mathrm{Martins}=\frac{37}{12}H,
    \label{eq:relationtoMartins}
\end{equation}
with $H$ defined in \eqref{eq:QISMhamiltonian}.

\section{Vicinity of the integrable point}\label{app:boost}

In this appendix we study the integrability in the vicinity of the model \eqref{eq:G2_integrable}. Specifically we consider the general Hamiltonian $H(\alpha,\beta)=\sum_iH_i$ with
\begin{equation}
H_i=\frac{1}{2}\mathbf{\Lambda}_i\cdot\mathbf{\Lambda}_{i+1}+\frac{\alpha}{4}(\mathbf{\Lambda}_i\cdot\mathbf{\Lambda}_{i+1})^2+\frac{\beta}{8}(\mathbf{\Lambda}_i\cdot\mathbf{\Lambda}_{i+1})^3
\label{eq:generalH}
\end{equation}
and periodic boundary conditions imposed. Grabowski and Mathieu~\cite{GrabowskiMathieu95} (see also Ref.~\onlinecite{GomborPozsgay21}) suggested a very hands-on way to check integrability of a given translationally invariant chain. They considered the boost operator 
\begin{equation}
B=\sum_{i=1}^N i\,H_i
\end{equation}
which yields a candidate for the first conserved charge
\begin{equation}
H_3=\bigl[B,H\bigr]=-\sum_{i=1}^N\big[H_i,H_{i+1}\big].
\end{equation}
If $H_3$ is indeed a conserved charge, i.e., $[H_3,H]=0$, then one must have
\begin{equation}
M_N\equiv\sum_{i=1}^N\Bigl[H_i+H_{i+1},\bigl[H_i,H_{i+1}\bigr]\Bigr]=0,
\end{equation}
which can be checked by straightforward calculation. This condition has to be satisfied for all chain lengths $N\ge 3$, in practice considering short chains is sufficient to get an idea. 

Applying the above argument to the model \eqref{eq:generalH}, we first calculate the eigenvalues of $M_3$. From this we deduce that the operator $H_3=[B,H]$ commutes with \eqref{eq:generalH} provided $\beta=15\alpha/37$ (alternative solutions are $\beta=-1+\alpha$ or $\beta=3(5\alpha-3)/31$, which we do not consider further). A constraint on the remaining parameter $\alpha$ is obtained from the requirement $M_4=0$, which is numerically found to be satisfied for $\alpha=5/2$ only, indicating that the general model \eqref{eq:generalH} is not integrable away from this point. 

\section{$(G_2)_1$ and Orbifolding}\label{app:Orbi}

As per the main text, our principal hypothesis for the low-energy, thermodynamic limit, properties of the $G_2$ integrable chain is that it is corresponds to a critical point described by a $(G_2)_1$ WZW CFT. Here we provide a short summary of the conformal data describing this field theory, and develop its orbifolded version, relevant for the analysis of conformal data via wavefunction overlaps.

\subsection{$(G_2)_1$ conformal data}
The $G_2$ exceptional algebra contains $14$ generators and has dual Coxeter number $g=4$, so that the conformal anomaly (central charge) of a $G_2$ WZW CFT at level-$k$ reads
\begin{equation}
    c_k=\frac{14 k}{k+g}\to c_1=\frac{14 }{5}= 2.8, \label{eq:cc}
\end{equation}
particularizing to our $(G_2)_1$ case of interest. 

The conformal dimensions of holomorphic primary operators are given by
\begin{equation}
    h_\lambda=\frac{C_\lambda}{2(k+g)}\to h_\lambda=\frac{C_\lambda}{10},
\end{equation}
again particularizing to $(G_2)_1$. $C_{\lambda}$ is the quadratic Casimir for a given $G_2$ irreducible representation (irrep). For level $k=1$, only two conformal towers exist, whose corresponding primary operators we name $\mathbbm{1}$, associated with the identity irrep, and $\tau$, associated with the fundamental, 7-dimensional, irrep of $G_2$. Their corresponding Casimirs read $C_{\mathbbm{1}}=0$ and $C_{\tau}=4$, so that their holomorphic conformal dimensions read
\begin{equation}
    h_\mathbbm{1}=0, \quad h_\tau=2/5.
\end{equation}

The scaling dimensions of non-holomorphic primary fields are related to the above by $h_{\lambda}=\Delta_{\lambda}/2$, thus 
\begin{equation}
    \Delta_\mathbbm{1}=0,\quad \Delta_\tau=4/5.
\end{equation}
These are the parameters that control the finite-size scaling of the amplitudes computed below.

The conformal characterization of $(G_2)_1$ is not complete without the fusion and modular content. The fusion relations for the primary fields are well-known and simple,
\begin{align}
    \mathbbm{1}\times\mathbbm{1}&=\mathbbm{1}\\
    \tau\times\mathbbm{1}&=\tau\\
    \tau\times\tau&=\mathbbm{1}+\tau,
\end{align}
following the same pattern expected for Fibonacci anyons. The quantum dimensions, which control the asymptotic Hilbert space size upon multiple fusions, are
\begin{equation}
    d_{\mathbbm{1}}=1,\quad     d_{\tau}=\frac{1+\sqrt{5}}{2}.
\end{equation}
We can now reconstruct the modular $\mathcal{T}$ and $\mathcal{S}$ matrices. For $\mathcal{T}$,
\begin{align}
    \mathcal{T}_{\eta\eta'}&\equiv \theta_{\eta}e^{-2\pi i\left(c/24\right)}\delta_{\eta\eta'},\ \ \theta_{\eta}\equiv e^{2\pi ih_{\eta}},\\
    \Rightarrow\mathcal{T}&=e^{-\pi i/45}\left(\begin{array}{cc}
1 & 0\\
0 & e^{4\pi i/5}
\end{array}\right),
\end{align}
where the Greek indices label primary fields. As for $\mathcal{S}$, the definition reads
\begin{align}
    \mathcal{S}_{\alpha\beta}&=\frac{1}{\mathcal{D}}\sum_{\eta}d_{\eta}C_{\alpha\beta\eta}\frac{\theta_{\eta}}{\theta_{\alpha}\theta_{\beta}}
\end{align}
where $C_{\alpha\beta\eta}$ is the fusion coefficient for three primaries and $\mathcal{D}=\sqrt{\sum_{\alpha}d_{\alpha}^{2}}$ is the total quantum dimension. To build the matrix easily, just note that the matrix must be symmetric and, since fusing with $\mathbbm{1}$ is trivial, the first row corresponds to the dimensions of the primary fields divided by the total quantum dimension. Then, unitarity demands the bottom right entry to be -1, and thus
\begin{align}
    \mathcal{S}=\frac{1}{\mathcal{D}}\left(\begin{array}{cc}
1 & d_{\tau}\\
d_{\tau} & -1
\end{array}\right).
\end{align}

\subsection{$(G_2)_1$ orbifolding}
The process of cyclic orbifolding a CFT is pedagogically explained in Ref.~\cite{Shinsei_orbifold}. Following their process, we quote the results for the $(G_2)_1$ orbifolding theory. For two primary fields in the parent CFT, five states are expected in the untwisted sector --- two symmetric, labelled `s', two anti-symmetric, labelled `a', and one mixed ---, and four states survive in the twisted sector. The twisted sector simply doubles the parent CFT primaries into new fields labelled with an extra index $\hat{0}$ or $\hat{1}$. 

Altogether, the primary-field content of the untwisted sector, and the corresponding scaling dimensions, read
\begin{align}
    \phi_{\left(\mathbbm{1},\mathbbm{1}\right)_{s}}&\to\Delta_{\left(\mathbbm{1},\mathbbm{1}\right)_{s}}=2\Delta_{\mathbbm{1}}=0, \nonumber \\
    \phi_{\left(\tau,\tau\right)_{s}}&\to\Delta_{\left(\tau,\tau\right)_{s}}=2\Delta_{\tau}=8/5, \nonumber \\
    \phi_{\left(\mathbbm{1},\mathbbm{1}\right)_{a}}&\to\Delta_{\left(\mathbbm{1},\mathbbm{1}\right)_{a}}=2\Delta_{\mathbbm{1}}+2=2, \nonumber \\
    \phi_{\left(\tau,\tau\right)_{a}}&\to\Delta_{\left(\tau,\tau\right)_{a}}=2\Delta_{\tau}+2=18/5, \nonumber \\
    \phi_{\left(\mathbbm{1},\tau\right)}&\to\Delta_{\left(\mathbbm{1},\tau\right)}=\Delta_{\mathbbm{1}}+\Delta_{\tau}=4/5.
\end{align}

For the twisted sector, we obtain
\begin{align}
    \phi_{\left(\mathbbm{1},\hat{0}\right)}&\to\Delta_{\left(\mathbbm{1},\hat{0}\right)}=c/8=\frac{1}{30}, \nonumber \\
    \phi_{\left(\mathbbm{1},\hat{1}\right)}&\to\Delta_{\left(\mathbbm{1},\hat{1}\right)}=c/8+1=\frac{31}{30}, \nonumber \\
    \phi_{\left(\tau,\hat{0}\right)}&\to\Delta_{\left(\tau,\hat{0}\right)}=c/8+\frac{2}{5}=\frac{13}{30}, \nonumber \\
    \phi_{\left(\tau,\hat{1}\right)}&\to\Delta_{\left(\tau,\hat{1}\right)}=c/8+\frac{2}{5}+1=\frac{43}{30}.
\end{align}

For our purposes here, we also need, at least part of, the fusion content of the orbifold theory. The selection rules for fusion of primaries $\alpha,\,\beta,\,\gamma$ are fixed by the integers  $\mathcal{N}_{\alpha,\beta,\gamma}$, typically assuming values $0$ or $1$. Whenever these integers assume a vanishing value, the full 3-point fusion can be discarded. The relevant coefficients for the orbifold theory can be obtained, using
\begin{align}
    \mathcal{N}_{\left(\alpha,\beta\right),\left(\gamma,\hat{\psi}\right),\left(\delta,\hat{\chi}\right)}=\sum_{\eta}\frac{\mathcal{S}_{\alpha\eta}\mathcal{S}_{\beta\eta}\mathcal{S}_{\gamma\eta}\mathcal{S}_{\eta\delta}^{*}}{\mathcal{S}_{\mathbbm{1}\eta}^{2}}
\end{align}
and 
\begin{align}
    \mathcal{N}_{\left(\alpha,\alpha\right)_{s},\left(\gamma,\hat{\psi}\right),\left(\delta,\hat{\chi}\right)}&=\frac{1}{2}\sum_{\eta}\frac{\mathcal{S}_{\alpha\eta}^{2}\mathcal{S}_{\gamma\eta}\mathcal{S}_{\eta\delta}^{*}}{\mathcal{S}_{\mathbbm{1}\eta}^{2}} \nonumber \\
    & +\frac{e^{i\pi\left(\psi+\chi\right)}}{2}\sum_{\eta}\frac{\mathcal{S}_{\alpha\eta}\mathcal{P}_{\gamma\eta}\mathcal{P}_{\eta\delta}^{*}}{\mathcal{S}_{\mathbbm{1}\eta}},
\end{align}
where $\mathcal{P}=\mathcal{T}^{1/2}\mathcal{S}\mathcal{T}^{2}\mathcal{S}\mathcal{T}^{1/2}$~\cite{Shinsei_orbifold}. Conveniently, all necessary orbifold fusion rules can be determined by the modular content of the parent CFT. Explicitly, the fusion rules of interest for the cyclic orbifold of $(G_2)_1$ read
\begin{align}
    \left(\mathbbm{1},\mathbbm{1}\right)_{s}\times\left(\mathbbm{1},\hat{0}\right)&=\left(\mathbbm{1},\hat{0}\right) \nonumber \\
    \left(\mathbbm{1},\mathbbm{1}\right)_{s}\times\left(\mathbbm{1},\hat{1}\right)&=\left(\mathbbm{1},\hat{1}\right) \nonumber \\
    \left(\mathbbm{1},\mathbbm{1}\right)_{s}\times\left(\tau,\hat{0}\right)&=\left(\tau,\hat{0}\right) \nonumber \\
    \left(\mathbbm{1},\mathbbm{1}\right)_{s}\times\left(\tau,\hat{1}\right)&=\left(\tau,\hat{1}\right)
\end{align}
starting from the symmetric identity sector, 
\begin{align}
    \left(\tau,\tau\right)_{s}\times\left(\mathbbm{1},\hat{0}\right)&=\left(\mathbbm{1},\hat{0}\right)+\left(\tau,\hat{1}\right) \nonumber \\
    \left(\tau,\tau\right)_{s}\times\left(\mathbbm{1},\hat{1}\right)&=\left(\mathbbm{1},\hat{1}\right)+\left(\tau,\hat{0}\right)  \nonumber \\
    \left(\tau,\tau\right)_{s}\times\left(\tau,\hat{0}\right)&=\left(\mathbbm{1},\hat{1}\right)+\left(\tau,\hat{0}\right)+\left(\tau,\hat{1}\right)  \nonumber \\
    \left(\tau,\tau\right)_{s}\times\left(\tau,\hat{1}\right)&=\left(\mathbbm{1},\hat{0}\right)+\left(\tau,\hat{0}\right)+\left(\tau,\hat{1}\right)
\end{align}
starting from the symmetric $\tau$ sector, and
\begin{align}
    \left(\mathbbm{1},\tau\right)\times\left(\mathbbm{1},\hat{0}\right)&=\left(\tau,\hat{0}\right)+\left(\tau,\hat{1}\right) \nonumber \\
    \left(\mathbbm{1},\tau\right)\times\left(\mathbbm{1},\hat{1}\right)&=\left(\tau,\hat{0}\right)+\left(\tau,\hat{1}\right)  \nonumber\\ 
    \left(\mathbbm{1},\tau\right)\times\left(\tau,\hat{0}\right)&=\left(\mathbbm{1},\hat{0}\right)+\left(\mathbbm{1},\hat{1}\right)+\left(\tau,\hat{0}\right)+\left(\tau,\hat{1}\right)  \nonumber\\ 
    \left(\mathbbm{1},\tau\right)\times\left(\tau,\hat{1}\right)&=\left(\mathbbm{1},\hat{0}\right)+\left(\mathbbm{1},\hat{1}\right)+\left(\tau,\hat{0}\right)+\left(\tau,\hat{1}\right).
\end{align}
for the mixed identity-$\tau$ one.

\subsection{Finite-size scaling and amplitudes}

In possession of the fusion rules for the cyclic orbifold version of a given CFT, one is ready to extract the conformal data from wavefunction overlaps. The general formula for the wavefunction overlap analysis described in the main text reads~\cite{Shinsei_orbifold}
\begin{align}
    A_{\alpha\beta\gamma}&=\left\langle \phi_{\gamma}^{3}|\phi_{\alpha}^{1}\phi_{\beta}^{2}\right\rangle \nonumber \\&=\sum_{\delta,\chi}a_{\left(\delta,\hat{\chi}\right)}N^{-\Delta_{\left(\delta,\hat{\chi}\right)}}C_{\left(\alpha,\beta\right),\left(\delta,\hat{\chi}\right),\left(\gamma,\hat{0}\right)},
\end{align}
where $a_{\left(\delta,\hat{\chi}\right)}$ are non-universal constants and $C_{\left(\alpha,\beta\right),\left(\delta,\hat{\chi}\right),\left(\gamma,\hat{0}\right)}$ are the operator product expansion coefficients for the cyclic orbifold CFT. Naturally, these coefficients are only finite when a fusion channel exists between the corresponding primaries.

It is convenient to normalize this expansion by the all-identity overlap $A_{\gamma\mathbbm{1}\mathbbm{1}}$, and noting that $C_{\left(\alpha,\beta\right),\left(\mathbbm{1},\hat{0}\right),\left(\gamma,\hat{0}\right)}=2^{-2\Delta_{\alpha}-2\Delta_{\beta}+\Delta_{\gamma}}C_{\alpha\beta\gamma}$ is related to the parent CFT operator product expansion coefficients $C_{\alpha,\beta,\gamma}$, we can write a general formula for the overlap of the $(G_2)_1$ CFT (in fact, valid for any WZW CFT at level $1$),

\begin{align}
    \frac{\left\langle \mathbb{\phi_{\gamma}}^{3}|\mathbb{\phi}_{\alpha}^{1}\phi_{\beta}^{2}\right\rangle }{\left\langle \mathbbm{1}^{3}|\mathbbm{1}^{1}\mathbbm{1}^{2}\right\rangle }&=2^{-2\Delta_{\alpha}-2\Delta_{\beta}+\Delta_{\gamma}}C_{\alpha,\beta,\gamma} \nonumber \\ 
    &+a'_{\left(\mathbbm{1},\hat{1}\right)}C_{\left(\alpha,\beta\right),\left(\mathbbm{1},\hat{1}\right),\left(\gamma,\hat{0}\right)}N^{-1} \nonumber \\ 
    &+a'_{\left(\tau,\hat{0}\right)}C_{\left(\alpha,\beta\right),\left(\tau,\hat{0}\right),\left(\gamma,\hat{0}\right)}N^{-\Delta_{\tau}/2} \nonumber \\ 
    &+a'_{\left(\tau,\hat{1}\right)}C_{\left(\alpha,\beta\right),\left(\tau,\hat{1}\right),\left(\gamma,\hat{0}\right)}N^{-\left(\Delta_{\tau}/2+1\right)}+....
\end{align}

The first term is the thermodynamic limit result, while the corrections arise at finite-size systems. Using the fusion channels from the previous section, we recover the three noteworthy overlaps from the main text, eqs.~\eqref{eq:A111}, \eqref{eq:A11gamma}, and~\eqref{eq:Agamma11}.

\section{Takhtajan--Babujian reference}\label{app:TKTB}

\begin{figure*}[h!]
    \centering
    \includegraphics[width=\textwidth]{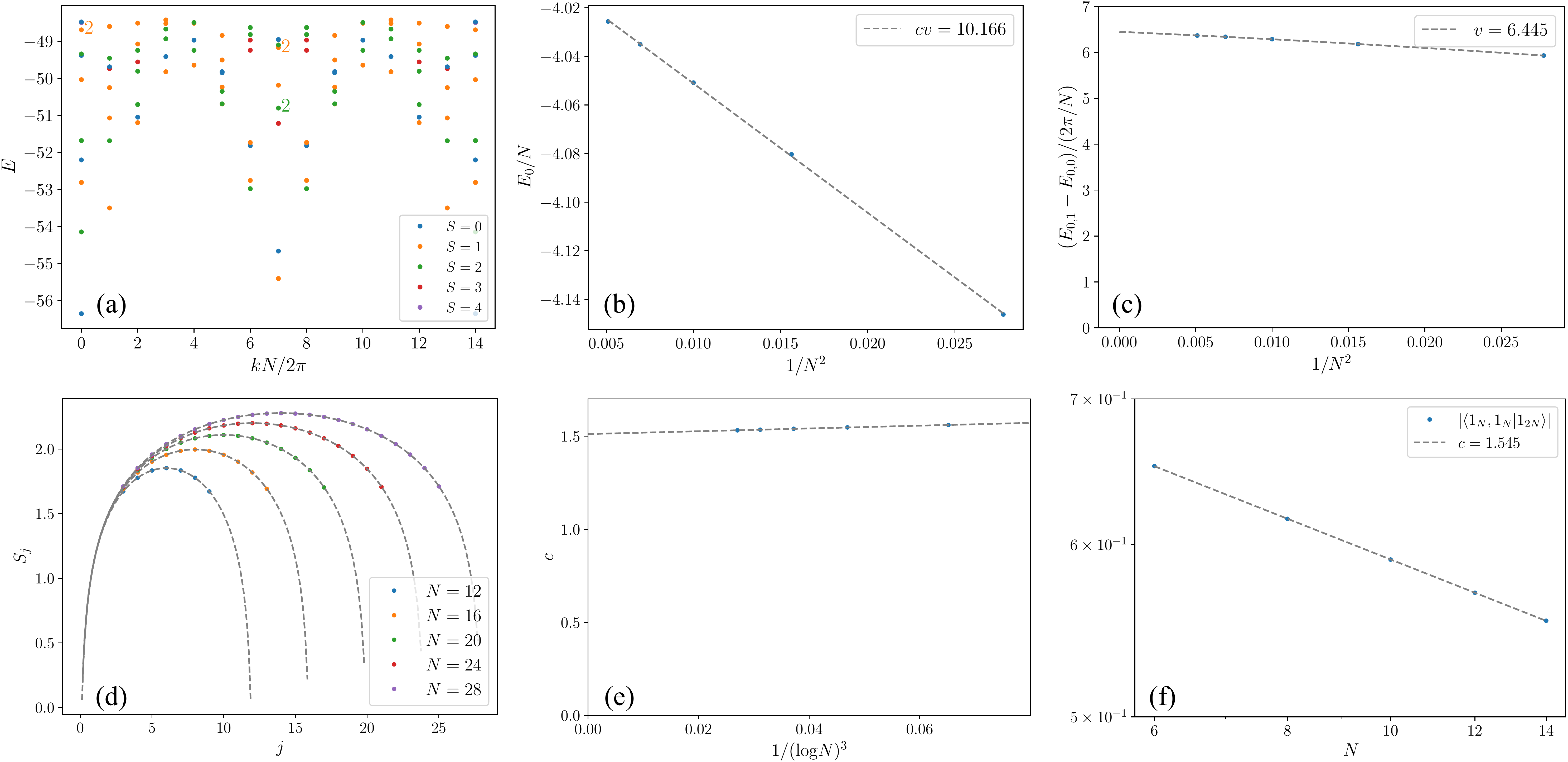}
    \caption{Benchmark results with the Takhtajan--Babujian model. (a) Momentum- and total-spin-resolved energy spectrum. (b) The scaling of the ground state energy. (c) Fitting of the spin wave velocity from $k=1$ and $k=0$ states, as done in the main text. (d) The entanglement entropy and central charge for various system sizes. (e) Fitting of $c(N)$ against $1/(\log N)^3$. (f) Wavefunction overlap and the resultant central charge. The extracted values of the central charge match $c\approx 3/2$, as expected.}
    \label{fig:tb}
\end{figure*}

In this Appendix, we present the numerical results on the TB model, serving as a benchmark for the three methods used in the main text for the $G_2$ model. The TB model is generally believed to be described by an $\mathrm{SU}(2)_2\sim\mathrm{SO}(3)_1$ WZW CFT. Both groups have dimension 3 and their dual Coxeter numbers are $g=2$ and $g=1$ respectively. This gives the central charge,
\begin{equation}
    c_k=\frac{3 k}{k+g}\to c_{\mathrm{TB}}=\frac{3}{2}= 1.5.
\end{equation}
We will see that all three methods give results consistent with this claim. 

The momentum- and total-spin-resolved energy spectrum is shown in Fig.~\ref{fig:tb}(a). The two-dome structure is consistent with that of $\mathrm{SU}(2)_2\sim\mathrm{SO}(3)_1$ WZW CFT. Extraction of the central charge from the energy spectrum constitutes determination of the spin wave velocity and a scaling of the ground state energy. The former gives $v=6.445$ (Fig.~\ref{fig:tb}(c)) and the latter gives $cv=10.166$ (Fig.~\ref{fig:tb}(b)), and together we have $c=1.58$, about 5\% off from the proclaimed value.

The entanglement entropy and the central charge for various system sizes are shown in Fig.~\ref{fig:tb}(d), with a fitting of $c(N)$ against $1/(\log N)^3$ in Fig.~\ref{fig:tb}(e). We see that even for such small system sizes the central charges are all rather close to $3/2$, with the thermodynamic limit value $c(\infty)=1.51$.

Finally, the wavefunction overlap data are shown in Fig.~\ref{fig:tb}(f). Due to difficulties in identifying the microscopic states with the CFT fields, we only consider the overlap among ground states, which directly gives the central charge. Once again the result is very close to $3/2$.

Before concluding this appendix, we wish to point out that in more careful analysis there turns out to exist subtle issues on numerical analyses and the identification of the TB model with $\mathrm{SU}(2)_2\sim\mathrm{SO}(3)_1$ WZW CFT. We refer the reader to Refs.~\cite{Alcaraz_1988,Affleck_TKTB} on this point.

\bibliography{spin3G2}

\end{document}